\documentclass[final,4p]{elsarticle}

\usepackage{cite}
\usepackage{url}
\usepackage{amsmath,amssymb,amsfonts}
\usepackage{listings}
\usepackage{algorithmic}
\usepackage{graphicx}
\usepackage{tabularx}
\usepackage{array}
\usepackage{footnote}
\makesavenoteenv{tabular}
\nopreprintlinetrue
\usepackage[caption=false,font=footnotesize,labelfont=sf,textfont=sf]{subfig}
\lstMakeShortInline[columns=fixed]|
\newcolumntype{P}[1]{>{\centering\arraybackslash}p{#1}}
\newcolumntype{M}[1]{>{\centering\arraybackslash}m{#1}}

\begin{document}

\begin{frontmatter}



\title{FAtiMA Toolkit - Toward an effective and accessible tool for the development of intelligent virtual agents and social robots}

\author{Samuel Mascarenhas, Manuel Guimar\~{a}es, Pedro A. Santos, , Rui Prada, Ana Paiva\\
INESC-ID \& Instituto Superior T\'{e}cnico\\
Universidade de Lisboa, Portual \\
samuel.mascarenhas@gaips.inesc-id.pt\\
manuel.m.guimaraes@tecnico.ulisboa.pt\\
rui.prada@tecnico.ulisboa.pt\\
pedro.santos@tecnico.ulisboa.pt\\
ana.paiva@inesc-id.pt}

\author{\\ Jo\~{a}o Dias, \\ INESC-ID \& Faculdade de Ciências e Tecnologia, Universidade do Algarve \& CCMAR}

\address{}

\begin{abstract}
More than a decade has passed since the development of FearNot!, an application designed to help children deal with bullying through role-playing with virtual characters. It was also the application that led to the creation of FAtiMA, an affective agent architecture for creating autonomous characters that can evoke empathic responses. In this paper, we describe FAtiMA Toolkit, a collection of open-source tools that is designed to help researchers, game developers and roboticists incorporate a computational model of emotion and decision-making in their work. The toolkit was developed with the goal of making FAtiMA more accessible, easier to incorporate into different projects and more flexible in its capabilities for human-agent interaction, based upon the experience gathered over the years across different virtual environments and human-robot interaction scenarios. As a result, this work makes several different contributions to the field of Agent-Based Architectures. More precisely, FAtiMA Toolkit's library based design allows developers to easily integrate it with other frameworks, its meta-cognitive model affords different internal reasoners and affective components and its explicit dialogue structure gives control to the author even within highly complex scenarios. 
To demonstrate the use of FAtiMA Toolkit, several different use cases where the toolkit was successfully applied are described and discussed.
\end{abstract}

\begin{keyword}
Embodied Agents, Affective Computing, Cognitive Architecture, Emotions, Social Robots


\end{keyword}

\end{frontmatter}

\section{Introduction}

FAtiMA Toolkit is a collection of tools designed to be used in tandem for the creation of cognitive agents with socio-emotional skills. The toolkit was developed in the context of the RAGE\footnote{\url{http://rageproject.eu/}} and Slice \footnote{\url{https://gaips.inesc-id.pt/slice/project.html}} projects with the goal of improving the FAtiMA Modular agent architecture \citep{dias2014fatima}. FAtiMA was initially developed in 2005 for the purpose of driving the behaviour of autonomous 3D characters in a serious game about bullying \citep{dias2005feeling} (the acronym stands for Fearnot AffecTIve Mind Architecture). At the time, its distinguishing feature was a computational model of an emotion theory named OCC \citep{ortony1990cognitive} that affected the character's planning and behavioural reactions. Since then, many other computational models of social and psychological constructs were added to the architecture. Some examples include empathy \citep{rodrigues2014process}, psychological drives \citep{lim2012creating} and culture \citep{Mascarenhas2016}. The addition of all these constructs had a significant impact on the complexity of using the architecture. To facilitate this problem, a modular version was developed \citep{dias2014fatima} that allowed users to turn off the cognitive modules they were not interested in using for their scenarios. Still, other usability issues remained in the modular version, which can also be found in other similar agent-based tools.

In a twenty year retrospective article about pedagogical agents \citep{johnson2016face}, Lewis and Johnson recognize that, despite the fact that there is plenty of empirical evidence of the effectiveness in the use of agents to promote learning, the lack of better tools for creating such agents is severely diminishing a more widespread adoption. Naturally, researchers have limited resources to spend on documenting, marketing, and supporting the tools they develop for doing research and this affects adoption by outsiders. However, there are more relevant and recurring problems in the design and implementation of agent-based tools that further aggravate this issue. Some of the problems are more of a technical nature such as the use of monolithic systems that do not favor heterogeneity of components, the framework paradigm, and how complex systems handle real-time interactions, but others can be more conceptual, such as how to explicitly structure the dialogue system (central vs distributed approach) or the lack of tools that facilitate the process of authoring characters and situations. These problems require a more flexible approach than the ones often used in cognitive agents.   

% Problems
% Rule Based Restrictions the limitation of a rule-based approach
% Framework Paradigm
% Rigid Dialogue Structures
%

In this document we discuss some of these problems, such as the ones mentioned before, and our proposed solutions that guided the creation of FAtiMA Toolkit. In particular, we will discuss the use of meta-cognition, which broadly speaking refers to the human capacity of ``thinking about thinking", as a solution. This capacity was represented by introducing meta-beliefs that are associated to dynamic reasoning processes, thus allowing for non-declarative knowledge to be interwoven together with declarative knowledge within a unified rule-based system. The added flexibility from this approach allowed us to tackle the identified problems, as described later in the paper. Our goal is that these solutions can serve as guidelines to other researchers that are developing similar systems.

To showcase the added flexibility and ease of application to different interactive environments and settings, we will describe different case studies where the FAtiMA Toolkit has been successfully applied. We will start with two commercial videogames that were developed by an external company using the toolkit: Space Modules, where players have to solve problems of customers that can be more or less angry and Sports Team Manager, where players most assemble the most optimally performing sailing team by resolving conflicts and managing the interactions between teammates. We also describe a Virtual Reality experience, a Game Jam involving FAtiMA and how a search algorithm such as Monte Carlo Tree Search was implemented in the Toolkit and applied to a survival game. Finally we will present some the results of using FAtiMA in a course on Game AI at IST, University of Lisbon, for the past 3 years and how we used the toolkit to investigate the impact of group-based emotions in a robotic partner.

%the use of the toolkit by 20 groups of students that had the task of creating short interactive stories with emotional characters as a course project. Finally, we present the use of the toolkit to investigate the impact of group-based emotions in a robotic partner. These case studies demonstrate that across heterogeneous contexts, technologies and requirements, different users were able to use the toolkit to create relatively complex interactive scenarios with socio-emotional characters.

In the remainder of this document we will start with a brief analysis of related work on computational models, frameworks, and tools for the development of socio-emotional agents. Next we will start by providing an overview of how FAtiMA Toolkit works along with its core components in Section \ref{sec:toolkit}. In the following two sections: \ref{contributions} and \ref{sec:dialog-structure} we will describe the main innovations of FAtiMA Toolkit. Agent-based frameworks have been found to create problems for their users and developers. For instance, having a monolithic structure for agent based development is quite inflexible and restrictive, hence we decided to use a library based design where each component is independent of the others. In these two sections \ref{contributions} and \ref{sec:dialog-structure} we discuss how we addressed these issues and the thought process behind our solutions. The first section describes the framework paradigm and our library based design and how the the introduction of meta-beliefs and reasoners allowed us to seamless integrate multiple cognitive processes. The latter sections delved into how and why we moved towards an explicit dialog structure that combines advantages of dialog tree structures with agent-based decision making.
The following Section \ref{sec:authoring} depicts the authoring tool created and its purpose. Finally the paper ends with the description and analysis of some of case studies that were created in past year with a  conclusion.

%Here we will start by  and how we addressed the problems that these architectures typically run into.  our solutions were developed by addressing t when designing this type of frameworks and the recently implemented solution of using a library based-design to address the problem of having a monolithic structure for agent based development. This section also describes how the introduction of meta-beliefs and reasoners allowed us to seamless integrate multiple cognitive processes. Section \ref{sec:dialog-structure} describes how we moved towards an explicit dialog structure that combines advantages of dialog tree structures with agent-based decision making. 

\section{Related Work}

As a cognitive agent architecture, the cornerstone of FAtiMA is its ability to synthesize emotions in a way that is grounded on human psychology with a particular focus on social interaction. Adopting a more physiological approach, the PSI model \citep{Bach06} establishes a cognitive motivation-based theory that aims at modeling the human psyche by formalizing its functioning in a computer simulation that regulates its needs autonomously. In PSI, all actions and behaviour ultimately have the purpose of achieving a finite set of primary, predefined drives. Instead of modeling explicit emotion categories, PSI sees emotions as emerging from specific modulations of cognitive and motivational processes. These modulations are realized by emotional parameters. Depending on the cognitive resources and the motivational state of the organism in a given situation, these parameters are adjusted, resulting in more or less careful or forceful ways of acting, as well as more or less deliberate cognitive processing. For instance, in the case of an event triggering an high arousal, it is likely that the system needs to react urgently to the situation at hand, which requires a fast decision process. PSI is an example of a physiologically inspired model, of which the Cathexis model \citep{Velasquez:1997} and the work of Cañamero \citep{Canamero:1997} are additional examples. Another type of a cognitive architecture that explores the interplay between several cognitive processes (e.g. reactive, deliberative) and sees emotions as mechanisms to regulate these processes is Sloman's CogAff framework \citep{Sloman:2002}. Although these type of models are very useful to understand and explain cognitive, behavioural and emotional processes, such complex and sometimes low-level descriptions of behaviour are not helpful to design socio-emotional characters for storytelling based scenarios, where simpler models are easier to understand and to use. 

EMA \citep{Gratch04,Marsella06,Gratch03} is an appraisal-based computational model that follows Smith and Lazarus's cognitive-motivational-emotive psychological theory \citep{Lazarus}, and extends SOAR \citep{Laird:2012} by incorporating emotions. In EMA, appraisal and coping are built on the causal representations developed for decision-theoretic planning augmented by explicit cognitive representations of desires, intentions, beliefs, the state of current plans and corresponding probabilities. Considering the example where a flying bird suddenly enters the room, EMA analyses the impact of this event in its current plans and goals. It discovers that there is a plausible hypothesis where the bird attacks and injures the agent. This possible future event creates a threat to an existing goal (being healthy), which will be considered undesirable and will then generate an emotion. Coping strategies are proposed to maintain desirable or overthrow undesirable in-focus emotion instances.

The WASABI architecture \citep{Becker-Asano+Wachsmuth.2010,Becker-Asano.2008} follows a dimensional approach where it simulates the whole agent's emotional state as a point in a continuous the three-dimensional space PAD (Pleasure, Arousal and Dominance). Pleasure determines the intrinsic contribution of an event to the agent's well being. Arousal represents the agent's level of being awake and reactive to stimuli. Finally, Dominance represents the agent's perceived capability to change the environment or the cause of the event causing a given emotion. Whenever an emotionally relevant event is perceived, the pleasantness of the event is appraised, generating an emotional impulse that will change the current level of pleasure and arousal in the PAD space, and which will slowly decay towards a neutral level. The level of Dominance is determined from the situational context from the cognitive layer of architecture. Primary emotions are used to trigger the corresponding facial expressions, while secondary emotions are sent to the deliberation component which can use this information to activate particular behaviours. We believe that this mechanism of making decisions based on emotional state information is important for an emotional agent architecture. However, while in WASABI the actual decision making-rules are tightly coupled to the implementation, in FAtiMA toolkit we adopted a declarative approach that allow authors to specify their own rules on how emotions affect behaviour, by using meta-beliefs about the agent's emotional state. 

The Virtual Human Toolkit \citep{hartholt_all_2013} is a well known architecture that is also designed to facilitate the creation of autonomous conversational characters. Moreover, the architecture is also highly modular. However, unlike FAtiMA Toolkit, the modules that are provided are focused in handling aspects that are more related to the embodiment of a character rather than its cognitive abilities. For example, the toolkit includes a multimodal sensing framework \citep{scherer_perception_2012} and a nonverbal behaviour generator \citep{lee_nonverbal_2006} that is based on the BML standard \citep{kopp2006towards}. Another distinguishing aspect is that the Virtual Human Toolkit adopts a framework approach that enforces the use of a message broker middleware for communicating with the available modules. 

GAMYGDALA \citep{popescu2014gamygdala} is a computational model of emotions that is based on the OCC theory\citep{clore2013psychological}. Similar to FAtiMA Toolkit, this emotional appraisal engine was designed to be more accessible to game developers. In the perspective of the authors, one of the issues with affective models that are tightly coupled to a specific reasoner is that their use is limited to games where that specific type of reasoning is desirable. With that in mind, GAMYDGALA implements an appraisal mechanism that is independent from the how the characters reason. Essentially, developers need only to provide a list of goals for each character and then specify which events will block or facilitate each goal. Based on that information, the engine will determine the changes made to the character's emotional state. Similar to GAMYGDALA, the appraisal engine in FAtiMA Toolkit is an independent component that is not tied to a particular type of reasoning. However, this is achieved in a different manner. Namely, developers are able to incorporate their game-specific AI as meta-beliefs, which can then be used by the appraisal process as regular beliefs.

The concept of meta-cognition, which refers to ``cognition about cognitive phenomena"\citep{flavell} is one of the inspirations behind CLARION.
 CLARION \citep{sun2006clarion} is an example of a modularly structured cognitive architecture that explicitly represents meta-cognitive mechanisms to enable a dual-process model of explicit and implicit cognition. In \citep{sun2007motivational} the author states that ``without meta-cognitive control, a model agent may be blindly single minded''. In order to promote a greater flexibility in its core cognitive algorithms the notion of meta-cognition was also integrated into FAtiMA Toolkit.

 %The notion of meta-beliefs that was integrated into FAtiMA Toolkit is based on the concept of meta-cognition, which refers to ``cognition about cognitive phenomena"\citep{flavell}.

%\section{Meta-Cognition in FAtiMA Components}
\section{FAtiMA Toolkit and its Core Elements}
\label{sec:toolkit}

\begin{figure}[h]
\centering
\includegraphics[width=0.68\textwidth]{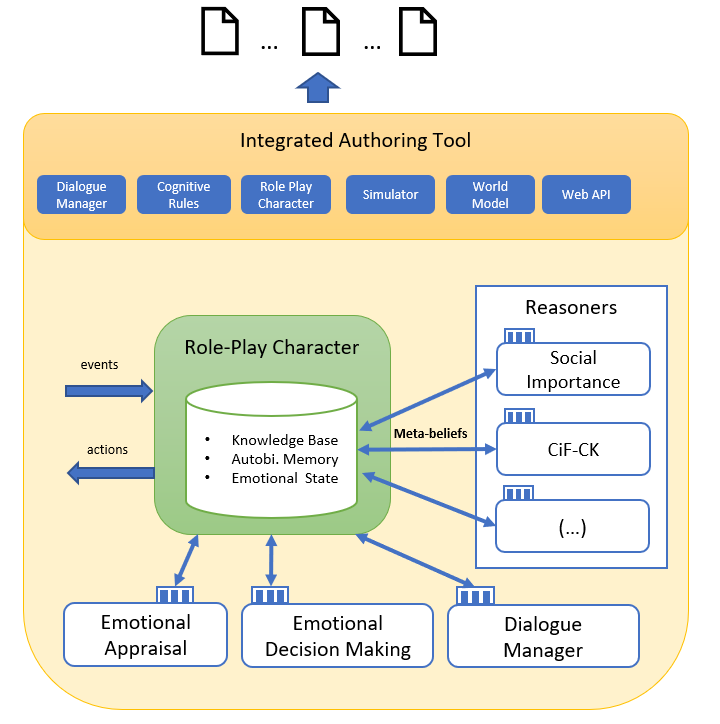}
\caption{\label{fig:fatima} FAtiMA Toolkit Components.}
\end{figure}

FAtiMA Toolkit is a collection of tools and assets designed for the creation of characters (virtual or robotic) with social and emotional intelligence. Figure \ref{fig:fatima} shows a diagram of all the existing components within FAtiMA Toolkit. Each is the result of continued iterations and experience accumulated while working with developers since the beginning of the RAGE project, now transitioning into the Slice Project. The entire toolkit is open source and is available on GitHub\footnote{https://github.com/GAIPS-INESC-ID/FAtiMA-Toolkit} and there are several resources, including Starter Kits, Demonstrations and Tutorials in FAtiMA Toolkit's official website \footnote{https://fatima-toolkit.eu/}.

An affective agent created with FAtiMA Toolkit will be initialized by the Role Play Character component. The name of the component was chosen based on the notion that FAtiMA agents are typically designed to be able to act according to a specific social role in a coherent manner. This component is responsible for managing the structures that constitute the internal state of the agent: their beliefs, memory and emotional state. Authors can also add more complex and believable behaviour to the agents, which we call Role Play Characters, by using other components. For instance, in order for agents to be able to perceive events in the world around them and have emotional reactions, the Emotional Appraisal component can be used. To define the agents' ability to make decisions based on their beliefs and goals one can use the Emotional Decision Making asset. All of these different components were developed as libraries, which the developers or authors can directly import and explore more easily without having to worry about compatibility issues. This is one of the main contributions of this work and we will further discuss it in Section \ref{contributions}. We will now describe the core elements that compose FAtiMA Toolkit

%The first one handles the agent's beliefs, the second one is responsible for storing and retrieving past events the agent witnessed and the third one manages the emotions the agent is experiencing and their decay over time. The Role Play character is also designed to integrate the functionality of the other components such as the Emotional Appraisal, Emotional Decision Making, Social Importance and CiF-CK. 

 \subsection{Events}

As typical of an agent model the main input are events. These can either refer to properties that change or to actions that occur. In both cases, events are represented as \textit{Well-Formed Names} (WFNs). These names are strings that follow a set of syntactic rules, which are based on the notions of symbols and variables from first-order logic:

\begin{itemize}
\item Symbols represent constant entities (actions, objects, name of properties, name of relations); e.g.: |Attack|, |John|, |Table|, |Brown|, |Loves|, |A1|, etc; There is one special symbol, |SELF|, which is always automatically replaced by the name of the agent.
\item Variables are written between square brackets and represent an entity or value that is not specified yet, which can be replaced by symbols; e.g.: |[x]|, |[target]|;
\item Composed names represent properties or relations between two or more Symbols; e.g.:\\ |Color(Sky, Blue)|, |Has(John, [x])|;
\\
\end{itemize}

As previously mentioned, events are represented as WFNs. More precisely, they are composed names that have the following form:

\begin{lstlisting}
Event(Action-End, Sam, Turn(On), TV)
Event(Property-Change, Sam, State(TV), On)
\end{lstlisting}

The first aforementioned example corresponds to an event where agent |Sam| finishes the action of turning on the TV and the follow-up example is an event indicating that the property |TV(Is)| changed its value to |On| due to agent |Sam|.

\subsection{Role Play Character}
\label{rpc}
The \textit{Role Play Character} handles the events in the following manner. If the event is a property change the corresponding belief in the KB is updated accordingly. If the event concerns the performance of an action, the agents checks if it recognizes the agent performing it. If not, the performer agent is added to the list of known agents. Subsequently, the event is processed by the \textit{Emotional Appraisal} component, which updates the emotional state of the agent with new emotions that the event might trigger. As will be detailed later, this process uses a rule-based approach. Finally, the event is stored in the \textit{Autobiographical Memory}, registering the event's logical name, the current time and the emotions it caused. 

The output of the \textit{Role Play Character} are actions or, more precisely, high-level decisions about which actions the agent should perform next. These actions can be read by external components upon request. The execution of these actions is left in charge of modules that are more closely linked to the agent's embodiment, such as a behaviour scheduler or a speech synthesizer. This type of modules fall outside the scope of FAtiMA Toolkit. To reach a decision, the \textit{Role Play Character} executes the algorithm defined in the \textit{Emotional Decision Making} and returns the result. Similar to the \textit{Emotional Appraisal} this algorithm also uses a rule-based approach. More precisely, it goes through the list of existing decision rules and applies, in each of them, a general unification algorithm between the list of the rule's activation conditions and the agent's set of beliefs. The algorithm will find all the possible sets of valid substitutions that satisfy the rule's conditions. Each of these substitution sets will originate a concrete decision to execute a certain action. To illustrate this process consider the following example of a decision rule:

\begin{lstlisting}
Action: Eat
Target: [object]
Priority: 1
Conditions:
  Has([object]) = True
  Edible([object]) = True
\end{lstlisting}

A single variable is present: |[object]|. As such, for this action to be executed a valid substitution must be found for it. Consider the following set of beliefs: ``|Has(Bag)=True|, |Has(Apple)=True|, |Edible(Apple)=True|". In this situation the unification algorithm would unify |[object]| with |Apple| but not with |Bag| as the latter fails to pass the last conditional check.

By expressing the agent's decisions using conditions with logical variables the action space of the agent grows and adapts according to its beliefs. This means that if an agent is able to find another object that it believes to be edible the previous rule would enable the action of eating it automatically. However, this type of rule-based reasoning by itself can be quite limiting when trying to handle scenarios that require a more complex decision-making strategy. For example, consider an agent that is intended to be able to play a game of go or poker in a proficient manner. In that case, it would be quite difficult to come up with decision rules based on logical belief checks. Because of this limitation, it is quite common to find other algorithms governing the decision-making of an agent, such as heuristic search, planning, reinforcement learning and deep learning. Any of these algorithms will be more advantageous in a particular domain and less so in others. For instance, deep learning has become immensely popular recently but it typically requires a large corpus of data to be useful. Also, the original version of FAtiMA included a classical continuous planner although it was difficult to explore its full capabilities in some conversational domains. 

\subsection{Knowledge Base}

Although the notion of beliefs and meta-beliefs has been already discussed, there are two important aspects that have not yet been mentioned. The first aspect is that the \textit{Knowledge Base} is capable of storing beliefs about the beliefs of other agents. In humans, this ability is referred to as a Theory of Mind. It was implemented here by specifying an additional field for each belief named |perspective|. The value of this field corresponds to the name of the agent that holds the belief. By default, the unification algorithm uses the agent's own perspective when searching for valid substitutions. However, this can be changed by using the following meta-belief in a condition: |ToM([agent],[belief])|. This meta-belief will retrieve the value of [belief] from the perspective of [agent]. Other meta-beliefs can also use this form of perspective-taking within their algorithm as is the case of the \textit{Emotional Appraisal} component when it is inferring the emotional state of other agents.

%rewrite this. uncertainty.
The final important aspect about beliefs is that they also have certainty value associated to them that can range from 0 to 1. The purpose of this value is to affect the perceived utility of each decision as will be explained later when describing the \textit{Emotional Decision Making} component.

\subsection{Emotional State}

The \textit{Emotional State} is the structure that holds the set of active emotions the agent is currently experiencing. When defining the nature of emotion, there were initially two main theoretical stances. The first of them considered that emotions could be grouped in several distinct categories or types \citep{ekman1971constants}. One of the issues with this approach is that their proponents have quite different views on how many categories exist and which criteria should be used to determine if a certain type of emotion should be considered as a primary type or not. This issue is exacerbated by the fact that some words used to describe emotions exist in some languages but not in others \citep{russell1983pancultural}. Motivated by these issues, a second theoretical stance proposes that emotions should be viewed as points in a n-dimensional space \citep{russell1980circumplex,russell1977evidence}. They circumvent the issues of the categorical approach by focusing only on broader characteristics of emotions that can be quantifiable such as arousal or pleasantness. Finally, a third view proposed that emotions are defined by the way we subjectively appraise our environment along several cognitive dimensions, also referred to as appraisal variables \citep{smith1985patterns,siemer2007same}.

FAtiMA is based on a combination of both the first and the third views, the OCC theory of emotions\citep{clore2013psychological}, is an appraisal theory that implicitly follows a categorical approach. Formally, an emotion is defined as a tuple with four elements: 
\begin{lstlisting}
<type,valence,intensity,cause,target>
\end{lstlisting}

The |type| of the emotion represents its categorical name. In total, the OCC theory proposes 22 emotion types that are psychologically distinct. Examples include Joy, Fear, Distress, Pride, Reproach, among others. The  |valence| of an emotion indicates if the emotion is positive or negative which depends on its type. For instance, Joy is considered to be positively valenced while Distress is viewed as negative. The |intensity| of an emotion is a number greater than zero that indicates how strongly is the agent experiencing the emotion. As times goes on, this value becomes smaller until it finally reaches zero and the emotion is removed from the emotional state. Finally, the |cause| corresponds to the event that triggered the emotion and the |target| is the name of the agent or object that the emotion is directed at.

One advantage that FAtiMA has in using a categorical approach to emotions, in comparison to a dimensional model, is that it is possible for agents to experience multiple emotions at the same time with different intensities. For example, the agent might experience Distress after losing its queen on a chess game while simultaneously feel Admiration for its opponent for doing a very good play. Another advantage is that it is also possible to specify different thresholds and decay rates for different emotion types. This can be used to simulate certain personality traits through the agent's emotional biases. For instance, using the previous example, an agent that is meant to be perceived as highly competitive in the chess example should be set with a lower threshold for Distress and a higher threshold for Admiration. 

Besides the list of active emotions, the emotional state also implements the |Mood([agent])| meta-belief that was previously mentioned. The concept of mood is an affective state that is more broad in the sense that it is not typically provoked by a specific event. Similar to emotions, mood is also characterized by a positive or negative valence but its duration is longer. Based on these notions, mood is represented in FAtiMA as a meta-belief whose value ranges from -10 to 10. This value is updated each time a new emotion is added. In the case of positive emotions, the mood value will increase based on the emotion's intensity. Conversely, negative emotions decrease the mood of the agent, also based on their intensity.

\subsection{Emotional Appraisal}
\label{EA}
The purpose of the \textit{Emotional Appraisal} library is to generate new emotions by doing a subjective assessment of the events received as input. The notion that emotions should be generated in this manner stems from the ideas put forward by appraisal theories of emotion \citep{moors2013appraisal}. On a broad sense, these theories propose that the reason why people have different emotional responses when facing the same situation is because they conceptualize it in a different manner. More specifically, people will make a series of judgments about a situation and the result of those judgments will determine the emotional response.

The appraisal model that is implemented inside the \textit{Emotional Appraisal} library uses appraisal rules to simulate the aforementioned psychological process of judging events. Formally, an appraisal rule has the following components:

\begin{lstlisting}
<event,target,appVariables,conditions>
\end{lstlisting}

The parameter |event| is a well-formed name that is used to match the name of the events received. The |target| is the name of the agent or object that the rule applies to. The third element, named |appVariables|, consists of a list of appraisal variables and their respective numerical values. Finally, |conditions| is a list of logical conditions that need to be verified according to the agent's beliefs in order for the rule to be activated. 

For the purpose of illustrating how the specified algorithm uses these rules to generate emotions consider the following simple example:

\begin{lstlisting}
Event: Event(Action-End,[x],Smile,SELF)
Target: SELF
AppVariables:
  Desirability = [d]
Conditions:
  RapportLevel(SELF, [x]) = [d]
\end{lstlisting}

Considering this rule, what would happen when an agent, named Sam, perceives the agent John smiling, which is represented as |Event(Action-End,Sam,Smile,John)|? Firstly, the algorithm tries to unify the rule's |event| with the event name. In this case, the unification is possible with the following substitution set \{|[x]/John|\}. It is important to mention that |SELF| is a special symbol that can be used in a rule and is always replaced by the name of the agent, in this case |Sam|. Because at least one substitution set was found, the next step is to unify the rules' |conditions| with the agents' beliefs taking into account the already existing substitutions from the previous step. If we assume that agent Sam believes |RapportLevel(Sam, John) = 5|, then the rule's |conditions| are made true by applying the following substitutions: \{|[x]/John, [d]/5|\}. The final step is now to generate an emotion based on the type and value of the appraisal variables specified in the rule. Following the OCC theory\citep{clore2013psychological}, a positive desirability results in the generation of a Joy emotion whose intensity is linearly proportional to the desirability value. Assuming this intensity is higher than Sam's threshold value of Joy, the emotion is added to the \textit{Emotional State} and becomes active until it intensity decays below a minimum value.

The appraisal process, specifically, is defined by two main steps, the Appraisal Derivation and the Affect Derivation. The first, Appraisal Derivation, determines what appraisal variables are affected by the event and their value. The second, Affect Derivation, computes what emotions are being triggered by those variables, their intensity and the resulting mood which in turn will affect the agent’s emotional state.

Authors have full control of the first process, the Appraisal Derivation, by defining Appraisal Rules and their components' values. The second process, the Affect Derivation, is internally computed by the framework and, as we mentioned, follows the model proposed by Ortony, Clore, and Collins \citep{clore2013psychological}. In order to support the full range of OCC emotions, FAtiMA uses 5 different appraisal variables:
\begin{itemize}
\item Desirability: Are the consequences of an event desirable for me?  
\item Desirability for others: Are the consequences of an event desirable for other(s)? Combined with the ``Desirability" appraisal variable describes emotions related to the Fortune of others and how desired that fortune is for the agent. 
\item Praiseworthiness: Is the action praiseworthy within the agent's social norms?
\item Goal Probability: How far is the agent from a reaching particular goal?
\item Like: Is the action/object/agent appealing?
\end{itemize}

The relation between these variables, their valence, and how each emotion is computed is briefly described in the OCC Model. During the implementation of those descriptions we've took some liberties as there was no exact equation on how, for instance, the intensity is calculated. Nevertheless FAtiMA is able to generate all 22 different emotions described in the OCC Model of Emotions. We can group these by the motivation behind them:

\begin{itemize}
\item Well-Being emotions: ``Joy" and ``Distress", described by the Desirability appraisal variable alone.
\item Fortune of others emotions: ``Happy-for", ``Gloating", ``Resentment" and ``Pity", are described by the combination of both Desirability and Desirability for others appraisal variables. Represent the consequences of an event towards others and how desirable that fortune is for the agent.
\item Attribution emotions: ``Pride", ``Shame", ``Admiration" and ``Reproach", described by Praiseworthiness alone, the main difference between them comes from the agent's responsibility. If the appraised action was performed others then it can lead to either “Admiration or Reproach” otherwise, if the agent was the ``executioner" it can lead to ``Pride" or ``Shame", depending on its valence.
\item Attraction emotions: ``Love" and ``Hate" are solely described by the Like appraisal variable. Describe the overall attractiveness of an object/event/action. 
\item Compound Emotions: ``Gratification", ``Remorse", ``Gratitude" and ``Anger", are the result of the combination between Desirability and Praiseworthiness appraisal variables.   
\item Prospect Based emotions: ``Hope", ``Fear", ``Satisfaction", ``Fears-confirmed", ``Relief" and ``Disappointment" are all related to events that contribute towards a possible future outcome, either positive or negative for the agent. Essentially they are described by the agent's Goals and how these evolve when an event is appraised. 
\end{itemize}

\subsubsection*{Goals and Goal Success Probability}

One of the main differences between the OCC Model and FAtiMA is how the framework handles prospect-based emotions and their relation with the agent's Goals. Here, we make two assumptions. First, we consider that all goals are something the agent wants to attain, as such they should all be ``positively oriented". For example, instead of having a ``don't die" goal authors should use “the agent wants to survive”.

The other assumption that we make is that the if the goal reaches extreme values, 0 or 1, they have reached a ``confirmed" status. If the likelihood of a goal is 0 then it has failed. If the likelihood of a goal is 1 then it has been achieved. Goals are defined by the following tuple:
\begin{lstlisting}
<goal,significance,likelihood>
\end{lstlisting}

The name serves to both identify and describe what is represents. The significance value describes how important the goal is to the agent, ranging 0 to 10. Finally, the likelihood parameter is the starting probability of the goal, as we mentioned before, ranges from 0 to 1. 

%% Falta falar de como este valor é updated

Prospect-based emotions are directly related to the evolution of the likelihood of each Goal. Goals-Confirmed emotions, Satisfaction and Fears-confirmed are only triggered if the Goal has reached either 1 or 0, respectively. The remaining 4 emotions are based on the expectation of the agent and if there was a major shift in the likelihood of reaching a goal. For example, if a goal such as “Survive” had a low likelihood of happening and, all of the sudden, its probability increases a lot then the agent feels Relief.

\begin{center}
  \begin{tabularx}{350pt}{  c | c | c }    
    \textbf{Previous Likelihood} & \textbf{Current Likelihood} & \textbf{Resulting Emotion }\\ \hline
    0.2 & 0.5 & Hope \\ \hline
    0.2 & 0.8 & Relief \\ \hline
    0.8 & 0.5 & Fear \\ \hline
    0.8 & 0.2 & Disappointment \\ \hline
    0.2 & 0 & Fears-Confirmed \\ \hline
    0.8 & 1 & Satisfaction \\ \hline
  \end{tabularx}
\end{center}

It is important to note that all ``unconfirmed" Prospect-emotions can be generated even if the goal is confirmed. For example, if the agent was expecting a goal to be achieved and it fails it will generate both emotions: “Disappointment” and “Fears-Confirmed”, however, the latter will have more intensity.

The process just described is also used to model the emotional state of other agents in the following manner. After the emotional state of the agent is updated based on the event perceived, the \textit{Role Play Character} component makes subsequent requests to the \textit{Emotional Appraisal} component to appraise the same event but from the perspective of the other agents. As a starting point, the agent assumes that all other agents have the same appraisal rules but the unification algorithm unifies the logical conditions of those rules using the beliefs it has of those agents' perspective and the keyword |SELF| corresponds to the name of the other agents instead. The resulting emotions are added to an identical emotional state structure which is created for each other agent that is present in the environment.

\subsection{Emotional Decision Making}
\label{EDM}
As previously mentioned, the purpose of the \textit{Emotional Decision Making} is to deliberate on the action that the agent should perform next. This is done by checking the list of the specified decision rules, which have the following attributes:

\begin{lstlisting}
<action,target,conditions,priority,layer>
\end{lstlisting}

%NOTE: Falar de layers como questão de deliberativa vs. reactiva%

The attribute |action| is a well-formed name that can be either a symbol (e.g. Smile, Eat) or a composed name (e.g. Offer(Gift)) in which the inner symbols are treated as the action's parameters. The |target| of the action is the name of the agent or object that the action is aimed at. The parameter |conditions| is a list of logical conditions used to check if the action can be executed. The |priority| is a numerical value that represents how important it is for the agent to execute the action. Finally, the |layer| parameter is a symbol that indicates in which layer is the action considered in the decision-making process. It allows the agent to consider different groups of actions at different time scales. Typically, there is a need for at least two layers when it comes to decision-making, one that is reactive and allows for fast decisions and another that is slower and focused on more strategic decisions. Note that the combination of these two layers is usually referred to as an hybrid agent architecture \citep{carrascosa2008hybrid}.

When describing the \textit{Role Play Character} and the notion of meta-beliefs we already used some examples of decision rules and how they work concerning the logical unification process. As such, we will now focus on the purpose of the last two attributes of a decision rule, namely |priority| and |layer|. Typically, there will be multiple decision rules in the chosen configuration of the \textit{Emotional Decision Making} component. Even if there is a single decision rule, it can generate multiple actions as a result of finding more than one set of valid substitutions. In that case, the agent needs to deliberate on which concrete action should it execute. This is what the |priority| of the decision rule is used for. More precisely, the \textit{Emotional Decision Making} will return the list of all actions found ordered based on their |priority| score, from highest to lowest. Note that the value of |priority| can be a logical variable that is determined by using both regular and meta-beliefs, similar to the appraisal variables in an appraisal rule. If that is the case, a numerical substitution must be found in the rule's |conditions| or the action is not selected. Alternatively, the value of |priority| can also be defined as a constant value. Finally, the |priority| score of a decision rule is affected negatively if the substitution set contains uncertain beliefs. This is done by multiplying the priority score with the certainty value of the beliefs that were used in the rule's substitutions.

One of the typical challenges associated with creating a decision-making solution for a virtual agent or a robot is being able to handle real-time interaction. This means that some of the agent's decisions must be made rather quickly in response to what is happening in the environment and what the other agents are doing. The purpose of the |layer| attribute is to solve this issue. Essentially, this attribute is used to group decision rules by their complexity and purpose. For instance, decision rules related to facial expressions can be labeled as `Expressive' whereas rules to decide upon which move to play in a given game can be labeled as `Deliberative'. Then, when making a decision request, it is possible to specify a specific |layer| as a parameter and the algorithm will only process decision rules of that |layer|. Without this mechanism, every decision rule would have to be analysed at each request, which would quickly become too inefficient for handling real-time interaction.

 \section{An accessible Intelligent Agent's Toolkit}
 \label{contributions}

FAtiMA Toolkit was specifically developed with the intent of making agent-based solutions more accessible and reliable for its users. Taking this into account we worked with several different authors, developers and students in order to identify the main problems when using this type of framework. In this section we will describe these issues and how we addressed them, we hope that our approach and solutions can serve as a guideline to other researchers that are developing similar systems.

%We will start by describing the motivation behind them and why our experience with working with d was quite important in their development. The problems described in this section require a more flexible approach than the ones often used in cognitive agents. 

\subsection{Library based design}
\label{library}
Many agent-based tools are developed by adopting a framework paradigm. More precisely, they define a rigid monolithic structure that acts as the skeleton of the whole system. For instance, in the case of FAtiMA Modular, the architecture dictates a core algorithm that implements the perception-action cycle of an agent. Within this core algorithm it is possible to add or remove components of different types. Due to this problem, the design of the core algorithm is faced with the highly difficult challenge of having to be generic enough to encompass all possible forms of a perception-action cycle in an agent. 

Another very important issue in the use of a framework paradigm is that, generally speaking, frameworks are hard to combine together due to their nature. Because of this issue, the risk in adopting a particular framework for a given project is substantially higher. In the context of the RAGE Project where we got to work with several different developers and received feedback from people within the industry we feel that this type of concern is quite important, specially in a large project. Additionally it plays also a major role in explaining why, for instance,  only a small number of game engines are actually used in the video game industry and many companies still decide to develop an in-house solution.

During the development of FAtiMA Toolkit we transitioned from a framework paradigm to a library based approach. Library based design means that the components work as individual libraries that can be used on their own. From a software standpoint this equates to while having a complex and robust architecture, the pieces that make part of it can function independently and can be picked apart to be used in another puzzle.  

In the FAtiMA Toolkit context is means that, if developers are only interested in having a solution to generate emotions they are able to use the \textit{Emotional Appraisal} library in isolation. If a developer is only interested in having a character that has an Autobiographic Memory then it can simply use the \textit{Role Play Character} library alone.

As we've mentioned, the components can be used autonomously, however they work best together. The configuration shown in Figure \ref{fig:fatima} depicts how the existing components are combined together to create an agent with emotions and a complete and rich perception-action cycle.

\subsection{Meta-beliefs}

%By expressing the agent's decisions using conditions with logical variables the action space of the agent grows and adapts according to its beliefs. This means that if an agent is able to find another object that it believes to be edible the previous rule would enable the action of eating it automatically. However, this type of rule-based reasoning by itself can be quite limiting when trying to handle scenarios that require a more complex decision-making strategy. For example, consider an agent that is intended to be able to play a game of go or poker in a proficient manner. In that case, it would be quite difficult to come up with decision rules based on logical belief checks. Because of this limitation, it is quite common to find other algorithms governing the decision-making of an agent, such as heuristic search, planning, reinforcement learning and deep learning. Any of these algorithms will be more advantageous in a particular domain and less so in others. For instance, deep learning has become immensely popular recently but it typically requires a large corpus of data to be useful. Also, the original version of FAtiMA included a classical continuous planner although it was difficult to explore its full capabilities in some conversational domains

On a practical level, the notion of meta-belief was implemented in the following manner. Similar to a regular belief, its syntactic structure is a composed name with a root symbol and a sequence of literals within parenthesis that are separated by commas. The difference is that meta-beliefs are registered in the Knowledge Base as procedures with their name becoming a reserved keyword. As such, the unification algorithm is able to identify them when parsing the composed names within a logical condition. Whenever it encounters a meta-belief it will dynamically execute its associated code to retrieve its value. The code can be either very simple or quite complex, depending on the nature of the meta-belief itself. For example, one of the meta-beliefs that the \textit{Role Play Character} adds is |Mood([a])|, which simply retrieves the mood value of agent [a]. Comparatively, the meta-beliefs added by the \textit{Social Importance} and the \textit{CiF-CK} modules are more complex. These two modules are labeled as \textit{Reasoners} in Figure 1 precisely because their main purpose is to augment the reasoning capabilities of the agent's decision-making and appraisal through the addition of meta-beliefs. The ``(...)" displayed in the diagram is to indicate that external developers can also extend the toolkit by adding additional reasoner modules with their own set of meta-beliefs without changing the toolkit's source code. Going back to the chess example, one possibility would be to add a chess engine as a reasoner module that would add the meta-belief ``BestChessMove([board-id],[player-color],[depth])". That meta-belief could then be used in a decision rule like this:

\begin{lstlisting}
Action: PlayChess
Target: [m]
Priority: 1
Conditions:
  Playing(Chess) = [id]
  PlayerColor(SELF,Chess) = [c] 
  BestChessMove([id],[c],Mood(SELF))=[m]
\end{lstlisting}

As this example illustrates, one of the advantages of meta-beliefs is that they can be used inside logical conditions in an identical manner as regular beliefs. Moreover, they can be nested inside each other. The nesting is possible because each meta-belief has full access to the contents of the \textit{Knowledge Base}, which includes both regular and other meta-beliefs. In this particular case, |Mood(SELF)| is nested inside |BestChessMove| with the aim of making the [depth] parameter of the chess search depend on the agent's current mood. The result would be that the agent, similar to a human, would come up with better moves in the game if it was in a good mood. To summarize, the key contribution of meta-beliefs, as defined in this manner, is that they allow for multiple reasoning and affective processes to be interwoven together under a unified rule-based system. We will now discuss in further detail some of the components that have previously been mentioned.

\subsection{Reasoner Components}
\label{Reasoner}
As stated previously, the role of reasoner components is to augment the capabilities of the \textit{Emotional Appraisal} and the \textit{Emotional Decision Making} by adding meta-beliefs that can be used in logical conditions. Therefore, their use is optional, which is important as they may require additional authorial effort. With that said, the two reasoners that are shown in Figure \ref{fig:fatima} work in conjunction to augment the ability of the agent to perceive and act appropriately in different socio-cultural contexts. More precisely, the first one, named \textit{Social Importance} was originally added to FAtiMA in order to be able to create groups of agents that would act and feel according to different cultural values \citep{mascarenhas2016modeling}. The second one, named \textit{CiF-CK} is based on a model \citep{MGuimaraes2017} that was used to add social behaviour to Non-Player Characters in the popular game Skyrim\citep{studios2015elder}. The model is an adaption of the ``Comme il Faut" architecture \citep{mccoy2010authoring}.
\\
\subsubsection*{Social Importance}
This component adds the meta-belief |SI([target])|, which calculates the amount of social importance (SI) the |[target]| agent has in the agent's perspective. This is a numerical value, ranging from 1 to 100, that signifies the extent to which the agent is willing to act in the interest of the |[target]| agent. The concept is based on a sociological theory of human motivation that conceptualizes human action in accordance to two fundamental relational dimensions: status and power \citep{Kemper2011}. According to this theory, we are motivated to confer status to those who deserve it and refrain from claiming more status than what we perceive to have in the eyes of others. The amount claimed and conferred by each action depends a great deal on cultural conventions but generally, the higher the request, the more status it claims. For a FAtiMA agent, the purpose of SI is to implement this notion of status so it can better navigate and understand different relational contexts. For instance, a stranger asking a personal question is inappropriate but not a friend. This is because, all else being equal, the SI of a friend should be higher than the SI of a stranger. These conventions are implemented in the component as attribution rules, which are defined as the following tuple:

\begin{lstlisting}
<target,conditions,siValue>
\end{lstlisting}

These rules work in a similar manner to the appraisal and decision rules described previously. Essentially, the unification algorithm will process each rule individually and try to find valid substitutions for the rule's conditions. If so, the |siValue| defined in the rule is added to the |target|'s total SI. Note that these rules can refer to beliefs about properties of other agents, such as whether or not they are a family member, or they can refer to their past actions in the environment, such as the amount of times they were rude towards the agent for example. Also, by using the meta-belief |ToM([target], SI(SELF))| the agent is also capable of inferring how much SI it has in the perspective of the |target| agent. This is done by using the beliefs associated to the |target| agent perspective when going through the attribution rules.
\\
\subsubsection*{CiF-CK} 
In most scenarios that are focused on social interaction it is likely there will be various social conventions that agents should follow such as introducing themselves in the beginning of a conversation or saying goodbye at the end. The purpose of the \textit{CiF-CK} component is precisely to make agents aware of this existing social exchanges as well as capable of executing them appropriately. The main motivation is to be able to automatically add a layer of social behaviour that can be reused across several domains.

The primary knowledge structure of CiF-CK are \textit{Social Exchanges}. These are meant to represent an established sequence of interaction steps between two agents. Formally, they are defined as the following tuple:

\begin{lstlisting}
<name, target, steps, conditions>
\end{lstlisting}

The parameter |name| is used to identify the type of social exchange (e.g. Flirt, Greet) and |target| is the name of the agent with whom the social exchange is to be performed with. The parameter |steps| is a list of symbols that represents the sequence of actions that both agents perform in a back and forth manner (e.g. \{|Initiate, Answer, Finalize|\}). Finally, |conditions| are logical conditions that can be of two types: (1) |starting-conditions| and (2) |mode-conditions|. The first type is used to define when a Social Exchange can start whereas conditions of the second type are used to compute the desire an agent has to perform a specific social exchange in a certain mode. As such, each |mode-condition| has the name of the mode it refers to (e.g. Positive, Negative, Sarcastic) and it also has a value associated with it. The desire for the agent to perform a step of the social exchange using a particular mode is given by the sum of the values of all its |mode-conditions| that are verified. 

The social exchanges that are defined in the \textit{CiF-CK} component can be used in decision rules through the following meta-belief: |SE([se-name],[target],[step],[mode])|. This meta-belief will automatically find valid substitutions for the defined variables and return a numerical value representing the desire of the agent to perform a specific step of a social exchange in a certain mode. For example, the value of |SE(Flirt, John, Initiate, Sarcastic)| is determined by first checking if the starting conditions of the ``Flirt" social exchange are verified and the current step in the execution order is ``Initiate". If all this applies, the algorithm then sums and returns the value of all the ``Sarcastic" conditions that are verified. The returned value can then be directly linked to the priority of the decision rule that uses this meta-belief. 

Note that there is a synergy between the \textit{CiF-CK} and \textit{Social Importance} component as it is possible to use conditions based on SI in the conditions of a social exchange. On the following section, we will now explain how a decision that involves dialogue such as the previous example of flirting is handled by the toolkit. 

\section{Explicit Dialogue Structure}
\label{sec:dialog-structure}

As previously mentioned, the first use case of FAtiMA was to drive the behaviour of non-player characters in FearNot!, a serious game about bullying. Rather than offering a linear story for the player to experience, the game chose to use an educational role-play approach, in which the story emerges from the interactions of characters playing their designated role (e.g. bully, victim, victim's friend, etc.). Like the player, each autonomous agent is controlling their designated character according to their own knowledge and perspective about the world and the other characters. Given that the dialogue in the story emerges from the individual decisions of the different agents, this approach to interactive storytelling is commonly referred to as being distributed or character-centric \citep{cavazza2002character}. Alternatively, the other common approach for designing an interactive storytelling system is the centralized or plan-based approach. In it there is a central AI component, typically referred to as a drama manager. This component is in charge of planning and directing all the dialogue actions that are made by each character and then adapting the story plan in accordance to how the player acts. The game Fa\c{c}ade \citep{mateas2003faccade} is a well-known interactive drama where this approach was used with success. There are both upsides and downsides in each approach. On one hand, the central planning approach provides a greater deal of control and predictability to the author but its rigidness makes it difficult to scale up. On the other hand, the distributed approach allows for much more flexibility in the way characters respond but loses in terms of predictability of the outcome, which also makes it hard to scale up when trying to assure a particular authorial or educational goal from the narrative.

Neither of the two approaches mentioned have been largely adopted in the video games industry so far due to its more risk-aversive nature. For the most part, developer studios still rely heavily on building scripted cut scenes mixed with multiple-choice options for the player. This usually leads to a branching narrative format. Consequentially, most game developers become accustomed to creating and managing dialogue trees, preferring to find ways by which their practical limitations can be compensated rather than switching to a distributed or centralized planning approach. Motivated by this preference, most commercial tools for building dialogue systems for video games also use dialogue trees as their principal component. 

\begin{figure}
\centering
\includegraphics[width=0.3\textwidth]{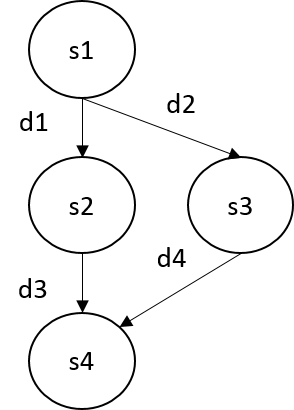}
\caption{\label{fig:dtree} Simple dialog tree structure with each circle representing a dialogue state and each arrow a dialogue action.}
\end{figure}

Because dialogues trees are so prevalent in game development, we decided to incorporate them in FAtiMA Toolkit but maintaining a character-centric paradigm. This lead to a hybrid solution that is meant to be more familiar to a game developer while still maintaining the flexibility and richness of having multiple autonomous agents controlling the dialogues of their characters. To explain how the proposed dialogue system works consider the tree structure shown in Figure \ref{fig:dtree}. The same structure could be represented in the following tabular format:

\begin{center}
  \begin{tabularx}{300pt}{  c | c | c | c }    
    \textbf{Id} & \textbf{Current State} & \textbf{Next State} & \textbf{Utterance} \\ \hline
    d1 & s1 & s2 & What are you doing? \\ \hline
    d2 & s1 & s3 & How are you feeling? \\ \hline
    d3 & s2 & s4 & Nothing special. \\ \hline
    d4 & s3 & s4 & I am feeling great. \\ \hline
  \end{tabularx}
  
\end{center}

Creating this simple dialogue between a player and a character would be quite straightforward in a system that is based on dialog trees. However, if one were to use an approach based on autonomous agents then the tree itself would no longer be explicitly represented anywhere. Instead, it would have to be implicitly defined in the agents' goals, beliefs, internal state, and actions. As such, to achieve the same effect, the author would have to configure an agent whose cognitive decisions would lead it to perform the dialogue action \textit{d3} after the player performed \textit{d1} and \textit{d4} after the player performed \textit{d2}. 

For such a simple dialogue, the agent-based solution is less practical. However, dialogues are usually quite less constrained than the example given. As one may notice, both in \textit{s1} and in \textit{s2} there is only one possible response for the character to select. As such, there is no need for a decision-making process. But what if we want to add multiple possible replies for the character to select from? For instance, consider that in state \textit{s3} we would like to have the following additional dialogue, to be said when the character is in a negative mood:

\begin{center}
  \begin{tabular}{  c | c | c | c }    
    \textbf{Id} & \textbf{Current State} & \textbf{Next State} & \textbf{Utterance} \\ \hline
    d5 & s3 & s4 & None of your business. \\ \hline
  \end{tabular}
\end{center}

In a system with dialogue trees, adding this dialogue for a character requires the existence of a conditional mechanism that will select it in the right context. In this case, an example would be a rule associated to \textit{s3} that would check the mood of the character and select \textit{d5} in case it was negative and \textit{d4} if it was positive. The scalability issue of this solution becomes apparent when we want a range of characters that exhibit diverse dialogue strategies in the same contexts. At this point, the rules encoded in each state must account for all the desirable character variations. This is the point where an agent-based solution becomes more valuable as each agent will have its own separate logic to select dialogue actions. However, due to this separation it becomes difficult to grasp how the narrative flows.

The solution adopted in FAtiMA Toolkit was to have an explicit dialogue tree structure but with a crucial distinction. The representation of the tree structure does not contain any logic that decides what the characters say. Instead, it merely informs agents controlling characters of how many options they can choose from in a given dialogue state. Essentially, it is as if the agents were other human players that can select freely among a finite set of dialogue options. In order for the agents to reason about their choice, dialogue options can be associated to different semantic tags pertaining to the meaning and the style of the utterance.

This information is accessible in the reasoning process of the agent through the following meta-belief:

\begin{lstlisting}
ValidDialogue([cs],[ns],[m],[s])
\end{lstlisting}

This meta-belief will go through the list of defined dialogues and will build a set of substitutions for each dialogue that it finds, also taking into account any existing constraints from previous conditions. The four different arguments are: [cs] - current state, [ns] - next state, [m] - meaning, [s] - style. For every substitution found, the value returned is `True'. To demonstrate the usefulness of this meta-belief consider the following decision rule for a `Speak' action:

\begin{lstlisting}
Action: Speak([cs],[ns],[m],[s])
Target: [x]
Priority: 1
Conditions:
  DialogueState([x]) = [cs]
  ValidDialogue([cs],[ns],[m],[s]) = True
\end{lstlisting}

This single decision rule is capable of handling any basic dialogue tree like the one shown in Figure \ref{fig:dtree} where the characters' dialogues are determined just by the current state of the conversation. Additionally, for the character to choose to speak using a `Rude' style when its mood is negative, adding the following decision rule with a higher priority would be sufficient:

\begin{lstlisting}
Action: Speak([cs],[ns],[m],Rude)
Target: [x]
Priority: 2
Conditions:
  DialogueState([x]) = [cs]
  ValidDialogue([cs],[ns],[m],Rude) = True
  Mood(SELF) < 0
\end{lstlisting}

The purpose of these examples is to demonstrate that, ultimately, the decisions are still being made autonomously according to the logic of each agent. As such, the flexibility that comes with an agent-based solution is not sacrificed. Instead, this flexibility is combined with a centralized representation of the possible dialogues, which is the paradigm that most game developers are already accustomed to. As a result, it becomes much less cumbersome for them to adopt this agent-based solution, even for scenarios with a simple dialogue structure. This centralized perspective for the definition of dialogues, meaning that they are not restricted in the mind of a single agent, is also aligned with the idea that dialogues are often social practices, of common knowledge, that can be used by different agents in similar situations.

\section{Integrated Authoring Tool}
\label{sec:authoring}
The accessibility of an agent-based framework like FAtiMA Toolkit is determined in large part by its authoring experience. Ideally, a person would be able to come up with different ideas for how a character behaves in one or more contexts and then easily configure different FAtiMA agents that would match those ideas without having to write code. 

\begin{figure}[h]
\centering
\includegraphics[width=0.7\textwidth]{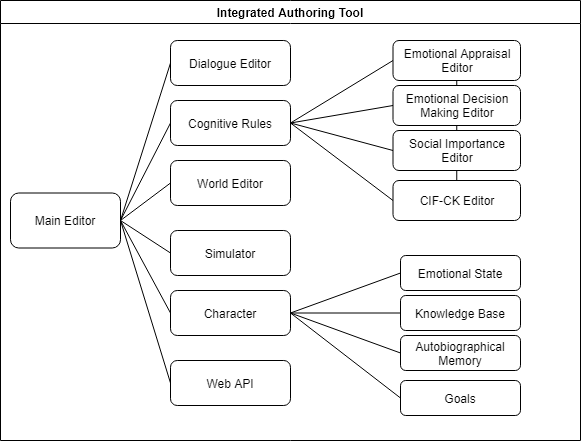}
\caption{\label{fig:iat} Integrated Authoring Tool - Diagram of existing components}
\end{figure}

With the previous goal in mind, FAtiMA Toolkit provides an integrated authoring tool that supports the creation and configuration of scenarios by non-coders. As shown in Figure \ref{fig:iat} this is a GUI tool with a hierarchical structure that contains several editors, a simulator and a Web API. At the top, there is a main editor where general aspects of the scenario are configured. These include the scenario's name, a short description and a list of character profiles. On the second level of the hierarchy, there are six different  editors with their own specific purpose. 

The purpose of the Dialogue Editor is to define all the dialogue actions that will be available for the agents and the players to select from. Additionally, there is also the possibility to export and import dialogues from other formats, including MS Excel. The Dialogue Editor also provides a validation mechanism that runs a depth-first search through all the dialogue states and tells the author which states are not reachable and how many end states exist. This information is used to quickly detect any possible errors in state transitions. Finally it is also possible to create and view a State Machine Graph of all possible dialogues and their corresponding states and transitions.  

The Character Editor is where the author configures the different agent profiles that are part of the scenario. For each agent, the author is able to specify their initial emotional state, the initial beliefs of their knowledge base, the previous events the agent will be able to recall in their autobiographic memory and the agent's goals. Each of these components has their own interface.

The Cognitive Rules section is responsible for configuring different cognitive components of the agents. In the Emotional Appraisal Editor, the author defines all the appraisal rules and parameters that drive the generation process of emotions. In the Emotional Decision Making Editor, the author specifies the rules that will guide the decision-making process of the agent. In the Social Importance Editor, the author specifies the rules that will tell the agent how it should perceive and act towards others from a relational perspective. Finally, in the CiF-CK Editor, the author can specify different patterns of micro-interactions that are common in human interaction such as a greeting or giving a compliment.

To help the authoring process, the author is capable of defining what are the consequences of an agent performing a specific action in the \textit{World Editor}. For instance, the author can specify that after each dialogue action the property ``Has(Floor)'' will assume the value corresponding to the name of the agent to whom the speak action was directed. This is particularly relevant in a multi-party conversation with multiple agents talking to one another. An additional rule could then be added that after a certain amount of time has passed if the agent who has the floor remains silent then the property ``Has(Floor)'' will assume the name of another random agent.

It is important to note that the effects of actions might be programmed directly in the game itself and communicated to the agents via events. The main benefit of using the World Editor is that these effects are configurable without having to recompile the game. Also, the action effects defined in the World Editor will be visible to the Simulator. 

The Simulator component allows the author to quickly run and test the defined scenario. As such, it becomes easier for the author to be able to do iterative testing and to reuse previous configurations from other scenarios and adapt them accordingly. Testers can play the role of any of the loaded characters or just become a spectator as the scenario unfolds. Additionally it also features several debug tools such as a real-time agent inspector capable of querying the agent's Knowledge Base, display their Emotional State, its Memory and Goals along with showing information regarding which rules are being triggered and its consequent effects. 

FAtiMA Toolkit's Web API module is our most recent component. Web APIs are a set of rules for interacting with a server with the most common use case being data retrieval. Essentially, it allows users to run a server on their own machine and through GET and POST functions interact with FAtiMA and its engine. 
The major benefit of Web APIs is, because they are built around HTTP protocols, nearly any programming language can be used to access them. As a result, FAtiMA is now able to be used by a wide array of different technologies and engines. 
For example, the video game Don't Starve Together is built on a custom C++ engine, however, most of the gameplay is written in LUA \citep{game:dontstarve}. This programming language supports HTTP communication, as such, it is possible to write a Mod for the game that communicates with FAtiMA and transforms the agent's decision rules within FAtiMA Toolkit into actions in the game. Fabio et al. did just that, they built a bridge between FAtiMA Toolkit and Don't Stave Together where authors can write decision and appraisal rules for an agent and watch the result of that authoring within the game, either cooperating with the agent or by simply watching as the action unfolds \citep{almeida}. We will further describe this particular Use Case in Section \ref{sec:usecase}.

%agent's are able to communiicate with FAtiMA through the Lua programming languge

One final consideration about the accessibility of the authoring process is that all of the configuration files are stored in a JSON format that is easy to read and change using a text editor.

\section{Use Cases\label{sec:usecase}}
\
To illustrate the potential and versatility of FAtiMA Toolkit, we will now present several different recent case studies where it has been successfully applied. We will start by enumerating each in the following Table and proceed to describe each more in-depth within the following sections.

%We will start by capturing and briefly describe most cases in a table and then and then we will then discuss a few of them more in-depth, namely, its FAtiMA's utility in Serious Games, Education and Social Group Dynamics with Robots. 

%The first one is a serious game that the company PlayGen developed with the help of the toolkit \citep{mascarenhasToolkit}. The second case study is the use of the toolkit by 20 groups of students that had the task of creating short interactive stories with emotional characters as a course project. Finally, we discuss our own use of the toolkit to drive the behaviour of a social robot in a team-based card game where the robot expresses group-based emotions.

\begin{center}
\label{usecases}
  \begin{tabular}{  m{4cm} | m{4cm} | m{2.5cm} | m{0.5cm} }  
  
    \textbf{Name} & \textbf{Project} & \textbf{Publication} & \textbf{URL} \\ \hline
    Space Modules 
    & RAGE Project 
    & \citep{mascarenhasToolkit, guimaraes2019accessible, westera2020artificial} 
    & \footnote{https://play.google.com/store/apps/details?id=com.PlayGen.SpaceModulesInc}
    \\ \hline
    Sports Team Manager 
    & RAGE Project 
    & \citep{mascarenhasToolkit, guimaraes2019accessible, westera2020artificial} 
    & \footnote{http://playgen.com/play/sport-team-manager/} 
    \\ \hline
    FAtiMA Virtual Reality Demo - Police Interrogation
    & RAGE Project
    & -
    & \footnote{https://fatima-toolkit.eu/showcase/} 
    \\ \hline
    FAtiMA's Game Jam 
    & RAGE Project 
    & \citep{guimaraes2019accessible} 
    & \footnote{https://itch.io/jam/fatima-jam}  
    \\ \hline
    Don't Starve Together 
    & Master Thesis 
    & \citep{almeida}
    & \footnote{https://fatima-toolkit.eu/fatima-toolkit-mcts-and-dont-starve-together/} 
    \\ \hline
    Game AI Course 
    & University of Lisbon 
    & \citep{guimaraes2019accessible} 
    & \footnote{https://fenix.tecnico.ulisboa.pt/cursos/meic-t/disciplina-curricular/845953938489420}
    \\ \hline
    Sueca Card Game With Robotic Partners
    & AMIGOS Project 
    & \citep{correia2018group}
    & \footnote{https://gaips.inesc-id.pt/amigos/}  
  \end{tabular}
%  \caption{\\ Table \ref{usecases}: FAtiMA Use Cases}
\end{center}

\subsection{Space Modules Inc.}

\begin{figure}[h]
\centering
\includegraphics[width=0.6\textwidth]{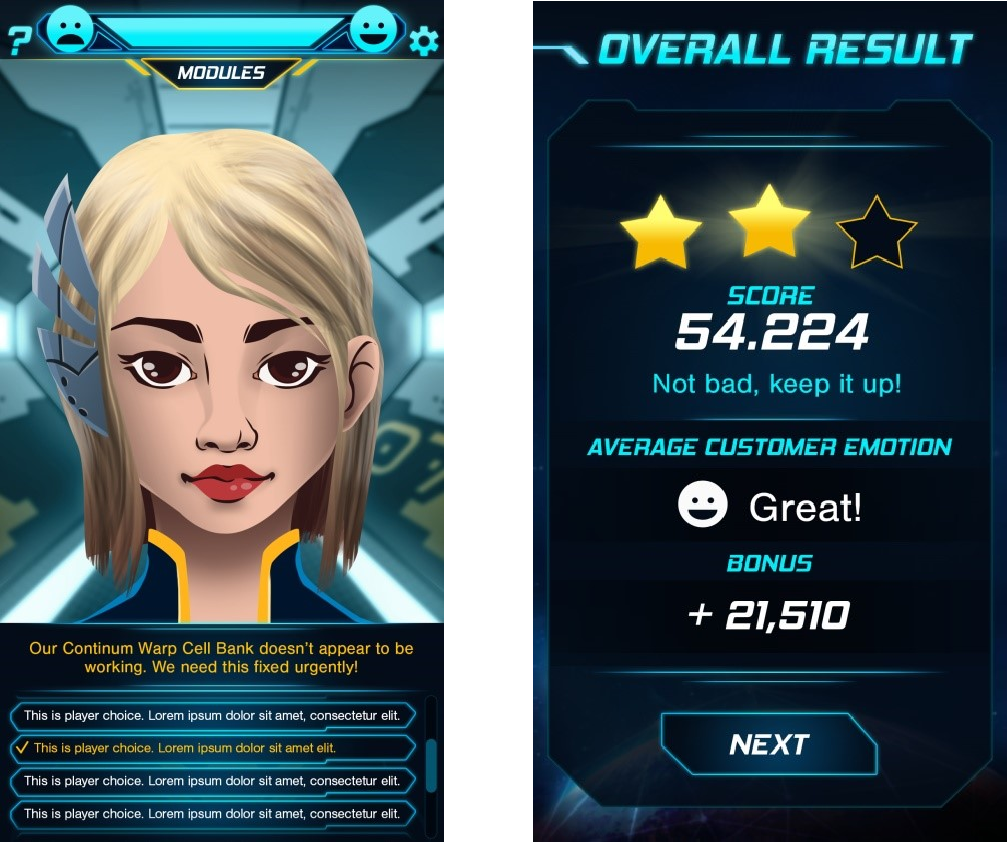}
\caption{\label{fig:space}SpaceModules - Conversation screen (left); Score screen (right)}
\end{figure}

%FALAR MAIS DO JOGo

Space Modules Inc (see Fig \ref{fig:space}) is a serious game that was developed in the RAGE project by the company PlayGen\footnote{http://playgen.com/} for an educational organization in the Netherlands named Stichting Praktijkleren\footnote{http://stichtingpraktijkleren.nl/}.  In total, the game will be used and evaluated across 6 participating schools with a total number of students around 1100. It is a single player game where the player must provide customer service on behalf of a spaceship part manufacturer. He or she does this by engaging in dialogue with different NPC customers that call the player to solve a problem they are having with their spaceship. The challenge lies in the fact that each customer has a unique emotional profile. For instance, some customers get angrier more quickly whereas other get angrier if you do not solve their problem promptly. The score that is attributed to the player is derived from the emotional state of the character at the end of the conversation.  Therefore, players must learn how to manage the emotions of their customers in addition to providing the right solutions to their problems. Concerning the interaction, a limited amount of dialog choices is given to the player at each turn until the conversation ends and a new customer call arrives. During the conversation, the NPC provides nonverbal feedback that is indicative of its current emotional state. It is up to player to pay close attention to these signals and the conversation unfolds. 

It is important to highlight that the integration of FAtiMA Toolkit in the code of Space Modules Inc. was handled directly by the game developers at PlayGen. By following an available example they were able to use the toolkit to drive the emotional dynamics of the characters as well as their responses. Moreover, the pedagogical experts were able to code the many dialogues that occur within the game using the explicit dialogue structure that was described in  section \ref{sec:dialog-structure}. This was achieved by first defining a list of symbols that would be used to categorize the  meaning and the style of the different dialogue alternatives. We then configured a small set of appraisal and decision rules that were sensitive to those symbols and to the game's pedagogical goals. After this was complete, the pedagogical experts were then able to use the provided authoring tool to code all the dialogues of the game taking into account the several different spaceship devices and their associated problems. This is a very encouraging result for the accessibility of the toolkit given that our involvement in the authorial process of previous serious games that used FAtiMA was much higher.

\subsection{Sports Team Manager}

Sports Team Manager is another game developed by Playgen during the RAGE project that makes use of FAtiMA Toolkit. Here, the objective is to assemble the most optimally performing sailing team by resolving conflicts and managing the team’s interactions. 

In order to better identify the character's skills and personalities, players must interview each member. The team has a set of predetermined roles, each with overlapping skill requirements \citep{mascarenhasToolkit}. Additionally players must take into account the team's morale and work on the social relationships within the team. Each customer/sailor can have a different emotional profiles, thus providing a different challenges and conflict situations, which the player must solve, will arise seamlessly.

\begin{figure}[h]
\centering
\includegraphics[width=1\textwidth]{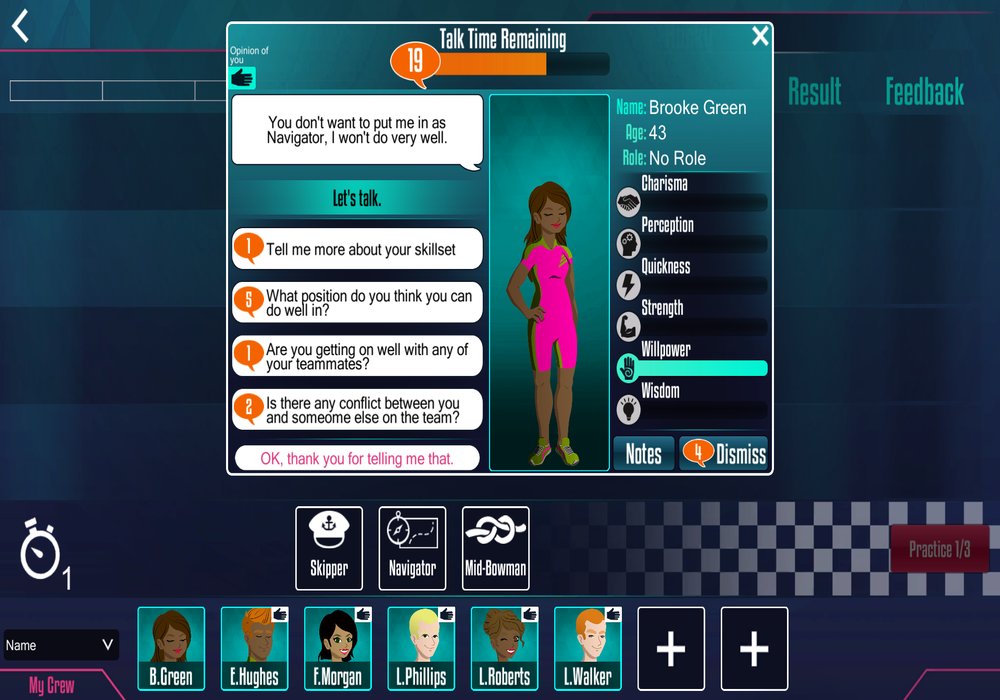}
\caption{\label{fig:sports}Sports Team Manager - Conversation screen}
\end{figure}

Just as it happened during the development of Space Modules Inc, the integration of FAtiMA with the game engine was handled directly by the developers. FAtiMA is responsible for managing the agent's beliefs, reactions and their actions. In this case, the beliefs are related to information such as their last position in the team, skill ratings, opinion ratings and event states. Sports Team Manager is also able to take full advantage of the Autobiographic Memory component of FAtiMA. Because each of the players decisions is saved into the characters memory, a history of events can be preserved. As such, it can be also be reloaded in further play sessions, allowing for the possibility of a persistent game emotional state and decision making. Figure \ref{fig:sports} shows an in-game screenshot of Sports Team Manager when a Player is talking directly to a particular character. 

\subsection{FAtiMA Virtual Reality Demo - Police Interrogation}

During the final years of the RAGE Project we had the opportunity of developing a Virtual Reality experience that makes full use of FAtiMA Toolkit's abilities. It also makes use of other RAGE-developed assets, such as L2F’s “Speech I/O” \footnote{https://www.gamecomponents.eu/content/226/speechio} and Behaviour Mark-up Language (BML) Realizer \footnote{https://www.gamecomponents.eu/content/231/bml-realizer}, showing players ,developers and other stakeholders, the synergy between the tools developed throughout the project. 

%In order to effectively design a place where players could properly interact with characters and scenarios built with the Storytelling Framework, we needed to set in place what type of interaction it would be and what type of relationship players would have with the characters. We chose to simulate a police interrogation. There are two main reasons behind this decision: First, one of the RAGE’s partners is Escola da Polícia Judiciária (Ministério da Justiça, PT), as such we could ask them for feedback and scenarios. Secondly, a police interrogation scenario can effectively test and assert the quality of our asset as it requires characters to display emotions and players to be focused on their decisions. 

\begin{figure}[h]
\centering
\includegraphics[width=1\textwidth]{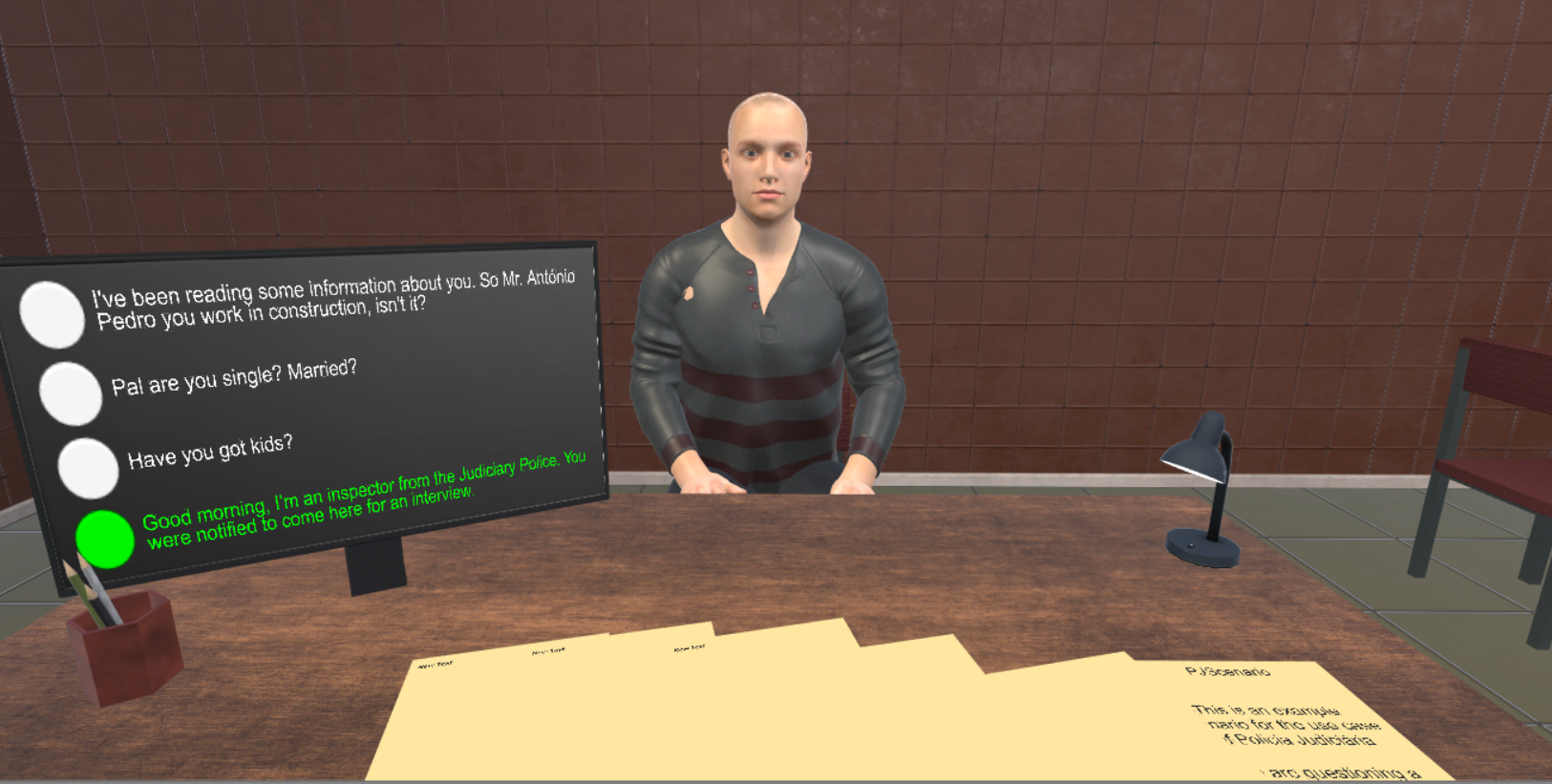}
\caption{\label{fig:police}FAtiMA Virtual Reality Demo}
\end{figure}

The demo is set in an interrogation room where users play the part of a  Police Interrogator and interact with a Suspect, that is a Role Play Character. The player’s objective is to obtain the most information about the suspect they can without losing control of the conversation. In order to interact with the subject, players can select questions from a wide array of possible choices that are displayed in a virtual computer screen as seen in Figure . After selecting a question the virtual suspect will answer to it, using Text to Speech, according to its mood and its emotions. The questions are sorted through various different topics such as “Marriage”, “Children”, “Work” among others. These questions and answers are managed through the Integrated Authoring Tool Asset’s Dialogue Manager. This experience, also called FAtiMA VR Demo, was developed using the Unity Game Engine and is compatible with most Virtual Reality headsets. Figure \ref{fig:police} shows the Player perspective within the game. The screen shown on the left is where the Players can take a look and select one of the possible questions to be asked to the Virtual Suspect.

\subsection{FAtiMA's Game Jam}

To promote the use of FAtiMA Toolkit an online ``hackathon" was created where participants were encouraged to create video games with engaging characters powered by the FAtiMA Toolkit. The hackathon was hosted on the website itch.io \footnote{https://itch.io/jam/fatima-jam} and lasted around 2 months.  

The winner was a game called ``The Princess and the Shapeshifter" \footnote{https://petzi.itch.io/the-princess-and-the-shapeshifter} that was created for the competition with original art and design. The developers took full advantage of FAtiMA's character-driven approach to allow players to take the role of the character's they interacted with. Throughout the game, players must interact with 6 different characters/agents, with different personalities and abilities and must persuade them into letting players take somethings from them so that they can "shape-shift" into them.

\subsection{Don't Starve Together}
\label{sec:dont}

Don’t Starve Together is an open world survival video game where, as in any survival video game, the objective is to surive for as long as possible \citep{game:dontstarve}. These games, typically, have several different non-playable-characters that players can interact with during their playthrough. The few autonomous characters of Don’t Starve Together present limited behaviour and are not subject to the same rules as the players, in terms of survivability. The lack of AI for NPCs capable of creating believable characters affects the player experience. This is the primary motivation behind Almeida's Master Thesis \citep{almeida}. 

In order to improve the quality the agents within the game, Almeida proposed the development of a platform that enables the creation of NPCs controlled by agency based models for survival games. Using FAtiMA Toolkit, as the model for agency, and Don't Starve Together, they've implemented and published the framework with an example character, Walter. Walter can be authored using FAtiMA Toolkit and its actions are reflected within the gameplay of a Don't Starve Together's session.

\subsection{Game AI Course}

The toolkit has been put to test in a course on Game AI at IST, University of Lisbon, for the past 3 years by over a hundred students. During the semester, students work on 4 different projects, the first three cover a wide range of fields within Artificial Intelligence: Movement and Collision Avoidance, Pathfinding and Monte Carlo Tree Search and Decision Making algorithms, correspondingly. For the fourth and final project students are given the option to choose the topic they want to delve into, within AI and Games. Here, they can propose their own assignment or they can choose one of the four different options we suggest, two of which use FAtiMA. One combines the FAtiMA's planning capabilities with Monte Carlo Tree Search applied to a ``survival" game and the other makes full use of the Toolkit's ability to create interesting conversational scenarios.  

%Often, when students are confronted with an open space of options they tend to get stuck in their own indecisiveness. Taking this into account we, the teachers, also provide different options of projects they might want to consider. Two of these proposed projects make use of FAtiMA. 

%While students can propose their own project we also provide Because it is such an open field and as a consequence there is a wide range of options we also suggest a number of different projects 

\subsection*{FAtiMA, Monte Carlo Tree Search and Don't Starve Together}

%In order to maintain the immersion created by the high levels of fidelity of graphics, in videogames, characters must be able to create the illusion of life, which requires them to possess basic human traits like social ability, reactivity, and active goal pursuit. Some game genres, like role playing games, have seen this problem being addressed by using agency based characters. However, survival games have not been subject to the same attention as the goal is often to just ``survive".

Since Almeida's Thesis \citep{almeida}, we've challenged Games AI students to improve the character (Walter), create their own model and try different approaches. For example, we've found that instead of creating a model with, let's say, 30 different decision making rules, a Monte Carlo Tree Search(MCTS) approach has better results in terms of survivability. In this case, students designed the agent's action space and their own interpretation of the environment around the agent and, using FAtiMA's Meta-Beliefs created their "MCTS" Algorithm within the framework. When agents used this meta-belief it would run a MCTS Algorithm and decide which of the available Don't Starve Together-based actions was the best according to their survivability goals. 

In terms of results, the standard ``survival" games' metric is number of days survived. The original ``Walter" survived for 1.8 days with 19 different decision rules. 2 years later students have been able to create an agent capable of surviving 5 days, on average. 

Over the past 2 years we've had over 20 different students in 10 groups, working on their own version of this project. Their results prove both the planning capabilities of FAtiMA and its adaptability to different contexts and environments.

%In this Almeida et al. work \citep{almeida}, they address this issue by proposing a framework that allows developers to create characters based on agency models for survival games. By making use of FAtiMA Toolkit, a fully fledged model for agency, and Don't Starve Together, a popular survival game, we've implemented and published such a framework with an example character, Walter. Walter has been run and tested against a behaviour tree based character. In 2018, during his Master Thesis, Almeida et al. focused on creating an agent-based framework for Don't Starve Together

\subsection*{FAtiMA Toolkit and Conversation Scenarios}

By opting into this project, students were given the task of creating at least two different conversation scenarios, one with a single character interacting with the player and another with two characters engaging in conversation with the player at the same time. While the project did not require students to work with all of FAtiMA's components, many of the groups choose to use most of them. Many of the students started by using the Emotional Appraisal and the Emotional Decision Making modules but, in order to enrich their scenarios they resorted to other reasoner components. Social Importance, for instance, was used to manage the friendship or romance between agents while the CIF-CK component was often used to accommodate social conventions such as agents greeting each other in the beginning of a conversation or saying goodbye at the end. In order to debug and test their scenarios each group used the Simulator and the Dialogue Editor quite a bit. One particularly useful and often shown in their reports was the option to create a graph tree with the authored dialogue states. 
%We give no restrictions regarding the theme of these scenarios, to increase diversity between projects. 

In order to get acquainted with the tool students, students had 1.5 hour workshops on FAtiMA Toolkit. Additionally we also shared with the students a version of the toolkit that is integrated with the Unity game engine and uses components to realize the body and expression of the characters developed by other members of the RAGE project.  

Over the last 3 years we've had over 30 groups of students (approximately 90 different people) creating and designing conversational scenarios using FAtiMA Toolkit for the course of AI on Games. Students were given no limitation, as long as the non-player characters had believable emotional responses they were free to use any kind of topic for the conversation. The result was a wide number of all kinds of surprisingly interesting scenarios. 

From a police officer pulling over the player, after-death experiences featuring ghosts, Obama and Trump debating about building a wall, to a scenario where players were at the gates of heaven and had to convince the gatekeeper to let them in. In the latter case, to be successful, players had to avoid upsetting the gatekeeper too much. Other groups opted for a more serious theme such as a job interview or a shopping scene with a father, his son, and a shopkeeper.

%On average, for each of the 37 scenarios, students wrote 80$\pm$38 different lines of dialogue distributed across 35$\pm$17 different dialogue states. Additionally each group created, around 7$\pm$4 different decision making rules and 6$\pm$2 appraisal rules, for each one of their Role Play Characters. Overall, this is an encouraging result as these diverse scenarios were created by students that had no previous experience with applying affective agents to interactive storytelling.

\subsection{Sueca Card Game With Robotic Partners}

\begin{figure}[h]
\centering
\includegraphics[width=0.6\textwidth]{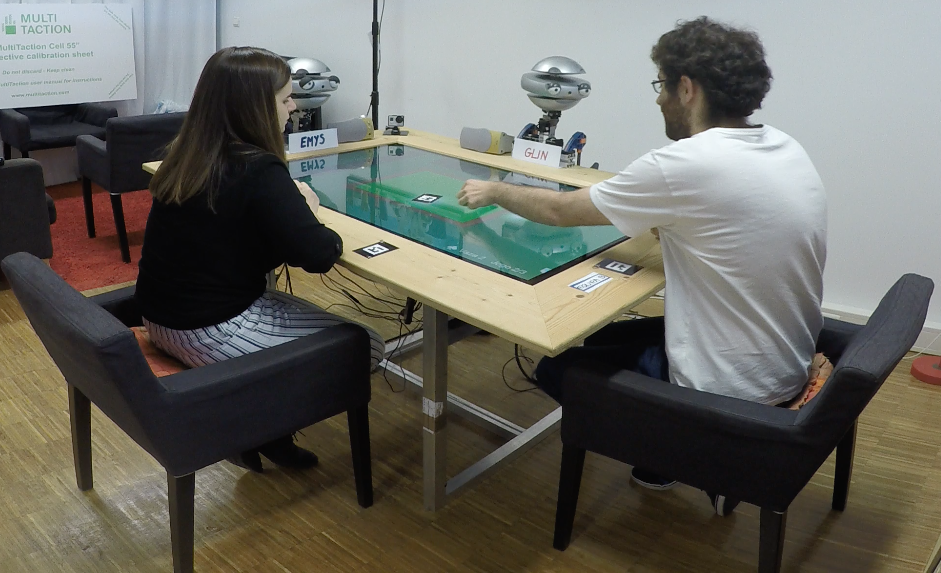}
\caption{\label{fig:sueca} Mixed human-robot teams playing Sueca}
\end{figure}

Our final case study involves the use of FAtiMA toolkit to control the decisions of social robots rather than virtual characters. There is also another important difference. Rather than focusing exclusively on dialogue actions, the case study revolves around a team-based card game. More precisely, two robots play the game of \textit{Sueca} with two human partners, as shown in Figure \ref{fig:sueca}. This is a Portuguese trick-taking card game that is played by four players divided in two teams against one another. 

From a research point of view, the main purpose of this game scenario was to investigate how robots can become more effective teammates. With that goal in mind, we decided to explore a particular type of affective phenomena that has not yet received a lot of attention in neither virtual agents nor social robots. The phenomena in question is group-based emotions, which is the term that is used to describe emotions that people experience due to their sense of membership to a given group \citep{kuppens2012group}. A common scenario where these emotions occur is when playing or watching competitive sports.

One of the interesting aspects of group-based emotions it that it combines appraisal theories of emotion with self-categorization theories of social identity \citep{turner1987rediscovering}. With this combination in mind, Goldenberg et al. \citep{goldenberg2016process} proposed a generative model of group-based emotions that extends the ``modal model of emotion'' by Gross \citep{gross2008emotion}. Based on the ideas of the model proposed by Goldenberg et. al., the \textit{Emotional Appraisal} component of FAtiMA was used to generate group based emotions by having the robot appraising both its actions as well as the actions of its partner as if both have the same subject, namely, their team. As an example, the robot felt and expressed ``group shame" whenever itself or its partner performed a bad play during the game. A user study was conducted where one of the robots in the game had group-based emotions and the other one felt individual-based emotions instead \citep{correia2018group}. The results show that the robot with group-based emotions was perceived as both more likeable and more trustworthy.

From a technical perspective, the two robots are fully autonomous in their behaviours. As detailed in \citep{AIIDE1715884}, a Perfect Information Monte-Carlo (PIMC) algorithm was used for the robot's ability to play the game in a proficient manner. The algorithm was combined with FAtiMA's appraisal mechanism so that emotions are generated based on how well the agent is doing in the game and are expressed through both utterances and non-verbal animations. In total, about 200 utterances were created for this scenario using the explicit dialogue structure that was defined previously. As for the non-verbal expressions, these are driven by Nutty Tracks, one of the components within the SERA ecosystem \citep{SSS1612748}. FAtiMA Toolkit's library design allowed the Emotional Appraisal component to be adapted to fit other frameworks and use different algortihms. Together with its follow-up study, over 100 participants interacted with the robots through FAtiMA designed scenarios \citep{correia2018group} \citep{AIIDE1715884}. 

%falar da capacidade para lidar com diferentes tipos de interacção

\section{Conclusions}

The importance of affective behaviour in characters (and agents) is widely acknowledged, in particular, in the gaming industry and in social robotics. Emotions capture our attention and make our interaction with artificial entities more natural and engaging. Consequentially, a lot of effort has been put into creating tools that facilitate the ability for these artificial entities to be able to portray different emotions and social behaviour with a increase degree of realism and believability. The use of affective models based on appraisal theories has the potential to automate the process of generating emotions for both virtual agents and social robots in a manner that is congruent with human psychology. Their application is particularly relevant when the social dynamics of the interaction are not trivial. This was the case of FearNot! \citep{dias2005feeling}, the serious game that led to the creation of the original FAtiMA architecture.

Despite their potential, the use of more complete affective models outside research has still been quite limited as there are both technical and conceptual issues that are hindering a more widespread adoption. One such issue that is on the technical side is the typical framework approach to agent models as it imposes limitations in how they can be applied and combined with other tools. On the conceptual side, cognitive architectures are designed with the intention of being able to function in many different domains. As such they need to be able to encompass different types of algorithms for decision-making and emotional appraisal. Finally, interaction scenarios with virtual agents and social robots tend to be heavily focused on dialogue but it is often the case that dialogue actions are defined as regular actions in an agent model without any further representation.

In this paper we described how these important issues were addressed in the development of FAtiMA Toolkit. Rather than using a framework approach, the toolkit uses libraries that are combined together to create affective agents. Moreover, through the use of meta-beliefs it is possible to use various specialized algorithms under a unified rule-based approach that governs both the agent's emotional appraisal and decision-making. The toolkit also has a unique approach to model dialogue that uses a centralized state-based structure for defining all the dialogue possibilities but leaving the decision logic in the agent model. Finally, a GUI-based authoring tool is also provided to facilitate the process of creating agent-based scenarios and testing them iteratively.

%The advantages brought by the toolkit were explored by a game studio that used it in the development of Space Modules Inc., a serious game for an educational institute. In this game, players give technical support to characters that act as angry customers. The toolkit was used to drive the characters' dialogues and emotional state, which is tightly coupled to the player's score. Previously, in other serious games that were developed with FAtiMA such as FearNot!\citep{dias2005feeling} or Traveller\citep{mascarenhas2013traveller}, the majority of the authoring work for the characters was carried out by us. This was not the case with Space Modules Inc., as all of the dialogue utterances and their logical representation was coded directly by pedagogical experts from the educational institute. To further test the toolkit, we asked students to use it in a project course to develop interactive stories with two characters talking with the player at the same time. In total, 20 groups of students that never used FAtiMA before were able to craft scenarios that had an average of 80 utterances. Finally, our last case study discussed how the toolkit was applied in a study of how group-based emotions impact the perception of a robotic partner in a card game. For this scenario, a specialized decision-making algorithm was integrated with the emotional appraisal provided by the toolkit.

The advantages brought by the toolkit have been explored by several different users in a number of different applications. In this article we presented a few of them. PlayGen, an UK-based game studio, used FAtiMA Toolkit in both of their games: Space Modules Inc. and Sports Team Manager. Players must deal with social dilemmas to accomplish their objectives and reach a better score. FAtiMA was used to manage these social interaction scenarios and to handle each of the character's behaviours. The FAtiMA Virtual Reality Demo - Police Interrogation is an experience whose focus is to display the high level of immersion a FAtiMA-modelled character could afford in a Virtual Reality setting. We have also illustrated the results of an Online Game Jam powered by FAtiMA and how a Master Thesis was able to implement a search algorithm through FAtiMA in a suvival game environment. Finally we also presented the outcomes of FAtiMA in a course on Game AI at IST, University of Lisbon, for the past 3 years and how we used the toolkit to investigate the impact of group-based emotions in a robotic partner.

%We have also described other FAtiMA-powered programs such as a Virtual Reality Police Interrogation serious game, an agent framework for applying a Monte Carlo Tree Search algorithm to a survival game and finally the results of an Online Game Jam. 

%We also described an experience whose focus was to display the high level of immersion a FAtiMA-modelled character could afford in a Virtual Reality setting: FAtiMA Virtual Reality Demo - Police Interrogation. 

%, a Game Jam involving FAtiMA and how a search algorithm such as Monte Carlo Tree Search was implemented in the Toolkit and applied to a survival game.  

Regarding future work, we are planning to provide additional tools to the toolkit with the goal of further alleviating the authoring effort, and improving control and verifiability of the socio-emotional behaviour of the designed characters. For instance, the hybrid dialog system design allows us to easily replace the user's role by an autonomous agent, which can be built from a particular user model, and then automatically test the dialog structure and decision making rules, determining the probability of reaching each particular state or expressing a particular behaviour. We believe this to be an important mechanism to provide game designers so that they can more easily detect potential problems or unreachable states and have some degree of authorial control over the interaction.

\section*{Acknowledgments}
This work was partially supported by national funds through  University of Lisbon and Instituto Superior Técnico and INESC-ID multi annual funding with reference UIDB/50021/2020 and the EC H2020 project RAGE http://www.rageproject.eu/; Grant agreement No 644187. 

\bibliographystyle{elsarticle-num}
\bibliography{main.bib}


%\newpage


%\vfill








\end{document}